\documentclass{article}
\usepackage{spconf,amsmath,graphicx,hyperref, tipa, longtable, booktabs, float}

\usepackage{comment}

\usepackage{caption}
\title{Validating Acoustic-to-Articulatory Inversion with Real-Time MRI: From Vocal-Tract Fidelity to Phonetic Consequences}
\name{
\shortstack{Tharinda Piyadasa$^{\star}$,
Michael Proctor$^{\dagger}$,
Amelia Gully$^{\ddagger}$,
Tuende Szalay$^{\star}$,\\
Emily Kiff$^{\dagger}$,
Kirrie Ballard$^{\star}$,
Craig Jin$^{\star}$}
}

\address{
$^{\star}$ University of Sydney, Sydney, Australia
$^{\dagger}$ Macquarie University, Sydney, Australia\\
$^{\ddagger}$ University of York, York, England
}

\begin{document}

\maketitle

\begin{abstract}
Acoustic-to-articulatory inversion provides a promising way to estimate vocal-tract behaviour directly from speech, but inferred articulatory variables need to be validated against independent physical measurements before they can be interpreted as faithful estimates of articulation. We investigate this using paired real-time MRI (rtMRI) and speech. A WavLM-based inversion model is trained to predict eight vocal tract variables (TVs) extracted from rtMRI, achieving a mean correlation of 0.643 on held-out speakers. Prediction accuracy is substantially higher for constriction degree than for constriction location, and the inferred trajectories systematically underestimate the dynamic range of rtMRI-derived articulation. We then use articulatory-to-acoustic synthesis as a functional probe of these errors, comparing speech generated from inversion-derived and rtMRI-derived TVs while holding the source features fixed. The discrepancy is small for vowels, approximants, stops, and nasals, but increases markedly for fricatives and affricates, where speech production depends on maintaining a narrow critical constriction and small geometric errors can strongly affect the acoustics. These results show that current acoustic-to-articulatory models recover broad vocal-tract behaviour well, but further work is required before inferred TVs can be regarded as faithful measurements of fine articulatory geometry.
\end{abstract}


\section{Introduction}
\label{sec:intro}

Acoustic-to-articulatory inversion seeks to recover vocal-tract behaviour directly from speech. Vocal tract variables (TVs) provide a compact, interpretable representation of constriction degree and location for the lips, tongue, and velum \cite{Saltzman89,browman1992articulatory}, and recent work has shown that they can be estimated using modern acoustic models and self-supervised speech representations \cite{siriwardena2023mirrornet,cho2024sparc,lee2025arti6}. Such representations support speech generation and articulatory coding, but may also be interpreted as estimates of physical speech production.

For this latter use, reconstructing speech accurately from an inferred representation is not sufficient. Acoustic-to-articulatory inversion is under-constrained: similar acoustics may arise from different vocal-tract configurations \cite{atal1978, Neiberg2008}, so an inferred trajectory may reconstruct speech well while still differing systematically from the articulation that produced it. A model may therefore produce smoother, compressed, or misplaced articulatory trajectories even when its acoustic output is plausible. Successful reconstruction does not establish that TVs faithfully measure vocal-tract behaviour. This distinction is particularly relevant to speech pathology, where articulatory information inferred from acoustics can inform assessment and rehabilitation of speech disorders involving altered motor or structural anatomy \cite{phippen2023}.

Real-time MRI (rtMRI) provides an independent way to test this question because it directly observes midsagittal vocal-tract geometry, including the tongue root, pharyngeal region, and velum \cite{wu2023deepmri,foley2025lss,lim2021multispeaker,tabatabaee2025nasality}. Recent multispeaker datasets and tracked contours enable comparison of acoustically inferred TVs with TVs derived from the imaged vocal tract \cite{lim2021multispeaker,shi2025annot16}. Previous work such as ARTI-6 \cite{lee2025arti6} demonstrates their utility for speech reconstruction and pseudo-labelling, but not their physical fidelity.

In this work, we pair rtMRI and speech to ask how faithfully acoustic-to-articulatory inversion recovers observed articulation, and where errors become acoustically consequential. A WavLM-based model \cite{chen2022wavlm} predicts eight TVs, which are compared directly with the rtMRI-derived trajectories. We then use articulatory-to-acoustic synthesis as a functional probe: a Transformer-based synthesizer \cite{vaswani2017attention} is driven by either inversion-derived or rtMRI-derived TVs from the same utterances while source information is held fixed. 

To our knowledge, previous work has not combined independent rtMRI-based validation with a matched synthesis comparison and phonetic analysis of the resulting errors. Our contributions are: (1) direct evaluation of inferred TVs against rtMRI-derived articulation; (2) a controlled synthesis-based test of whether the resulting discrepancies matter acoustically; and (3) phonetic analysis identifying the articulatory configurations where errors are most consequential.

\section{Methods}
\label{sec:method}

\begin{figure*}[t]
    \centering
    \includegraphics[width=\textwidth]{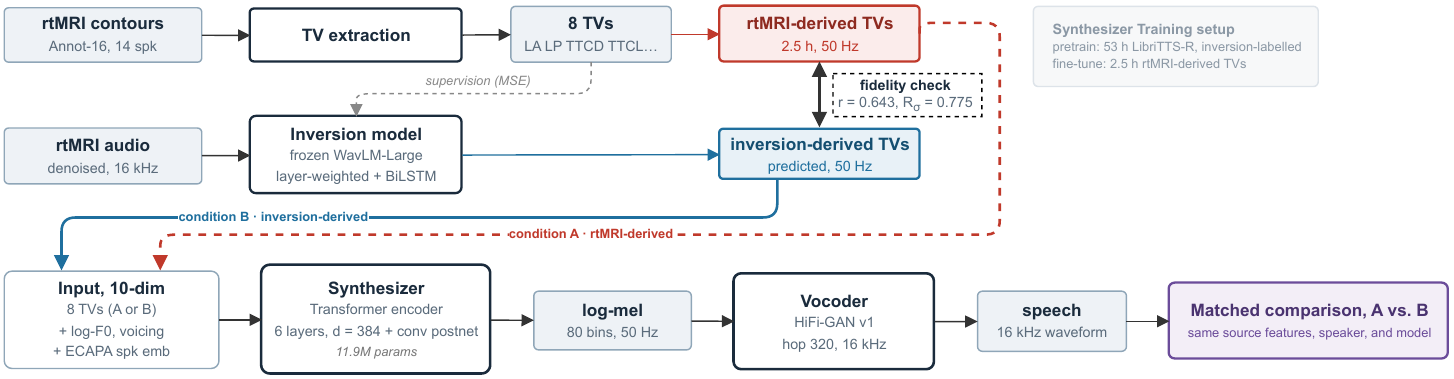}
    \caption{Experimental framework. Tract variables are obtained in two ways from the same rtMRI recordings: directly quantified from tracked contours (rtMRI-derived) and predicted from the audio by a WavLM-based inversion model (inversion-derived). The two are compared directly, then used to drive the same synthesizer on the same held-out utterances, with source features and speaker conditioning held fixed, so that differences in the output isolate the effect of TV provenance. The synthesizer is pretrained on inversion-labelled LibriTTS-R and fine-tuned on the rtMRI corpus.}
    \label{fig:pipeline}
\end{figure*}

Fig.~\ref{fig:pipeline} summarizes the experimental framework. Our objective is to assess how faithfully TVs inferred from acoustics represent articulation observed independently with rtMRI, and to determine when discrepancies between the two representations become acoustically consequential.

\subsection{Data and rtMRI-derived tract variables}
\label{sec:tvs}

We use 358 utterances from 14 speakers of the Annot-16 corpus \cite{shi2025annot16}, which provides tracked articulator contours together with phone- and word-level forced alignments for the USC 75-speaker rtMRI database \cite{lim2021multispeaker}. The material includes read passages, nonsense VCV
sequences, spontaneous picture descriptions, and topic monologues. Audio was recorded in-scanner at 16\,kHz and denoised using a waveform-domain enhancement
model \cite{defossez2020realtimespeechenhancement}.

\begin{figure}[t]
\centering
\includegraphics[width=\columnwidth]{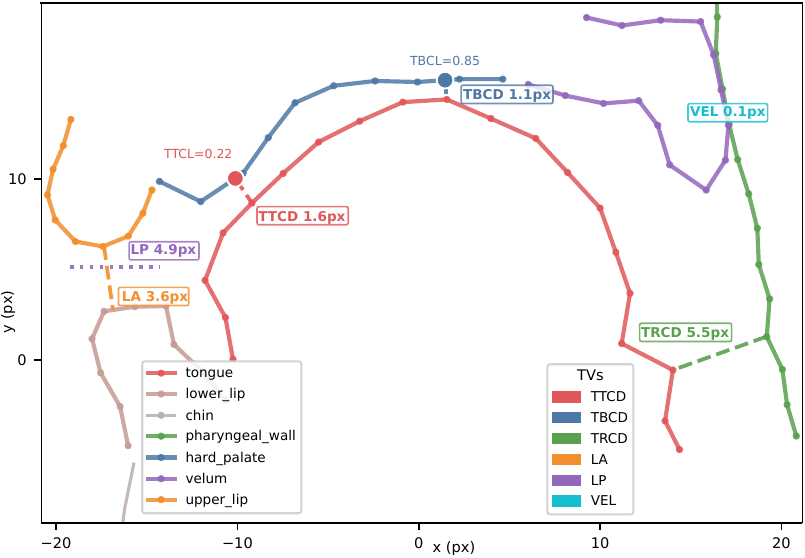}
\caption{Extraction of the eight rtMRI-derived tract variables from the tracked midsagittal contours of a single rtMRI frame.}
\label{fig:tvs}
\end{figure}

We derive eight TVs from the tracked midsagittal contours (Fig. \ref{fig:tvs}): lip aperture and protrusion (LA, LP); tongue-tip, tongue-body, and tongue-root constriction degrees (TTCD, TBCD, TRCD); tongue-tip and tongue-body constriction locations (TTCL, TBCL); and velic aperture (VEL). The TVs extend those of Articulatory Phonology \cite{Saltzman89,browman1992articulatory} to exploit structures visible in rtMRI: pharyngeal constriction is represented by a dedicated tongue-root constriction degree (TRCD) \cite{tilsen16}, which rtMRI resolves directly.

Constriction degree is the minimum Euclidean distance between a tongue region and its target passive articulator: the hard palate for TTCD and TBCD, and the posterior pharyngeal wall for TRCD. The velum is excluded from the tongue target surfaces because it is itself a mobile articulator with its own TV, and including it would couple TBCD to velic lowering under a static tongue. Constriction location is normalized arc length along the hard palate, from 0 at the alveolar ridge to 1 at the velar junction, so TBCL saturates at 1 for tongue-body configurations posterior to the hard palate. Tongue regions are delimited along normalized tongue-contour arc length from the tip as tip ($\leq 0.45$), body ($\geq 0.60$), and root (posterior $0.35$). LA is the minimum inter-lip distance, LP the protrusion of the lip midpoint relative to the anterior palate, and VEL the minimum velum-to-pharyngeal-wall distance. As rtMRI resolves the velopharyngeal port directly, we retain a continuous aperture measure \cite{tabatabaee2025nasality}. We refer to these throughout as \emph{rtMRI-derived TVs}, to distinguish them from variables predicted from speech acoustics. The underlying contours are themselves estimates: Annot-16 reports a 3.58-pixel Hausdorff distance between semi-automatic and expert handmade tongue contours \cite{shi2025annot16}.

\subsection{Acoustic-to-articulatory inversion}
\label{sec:inversion}

The acoustic-to-articulatory model predicts the eight TVs from speech. We use a frozen WavLM-Large encoder \cite{chen2022wavlm}, combine its hidden representations using a learned layer weighting, and pass the resulting features through a projection and a two-layer bidirectional LSTM with 256 units per direction. A final linear layer predicts the eight TVs at the 50\,Hz WavLM frame rate. The rtMRI-derived trajectories are resampled to the same frame rate. The model is trained using a mean squared error (MSE) objective on approximately two hours of paired rtMRI--speech data. Checkpoints are selected using mean Pearson correlation, and the evaluation is speaker-independent.

\subsection{Assessing articulatory fidelity}
\label{sec:fidelity}

Correlation alone does not establish that an inferred TV preserves the range of the underlying articulatory trajectory. For rtMRI-derived trajectory $z$ and inversion estimate $\hat z$, we therefore report Pearson correlation $r=\mathrm{corr}(z,\hat z)$ and the predicted-to-rtMRI standard-deviation ratio $R_\sigma=\sigma_{\hat z}/\sigma_z$. While $r$ measures how well the inferred trajectory follows articulatory variation, $R_\sigma<1$ indicates compression of the rtMRI-derived articulatory range.

\subsection{Synthesis as a functional probe}
\label{sec:synthesis}

Direct comparison of $z$ and $\hat z$ establishes whether the representations differ geometrically, but does not establish whether those differences matter for speech production. We therefore use articulatory-to-acoustic synthesis as a controlled functional probe.

Because the paired rtMRI corpus is too small to train a high-capacity synthesizer reliably from random initialization, the inversion model is first applied to LibriTTS-R \emph{train-clean-100} \cite{koizumi2023libritts}, producing 31,651 pseudo-labelled utterances totalling 53.0\,h. For each utterance, we extract the eight inversion-derived TVs, log-$F_0$, continuous voicing probability, and an ECAPA-TDNN speaker embedding \cite{Desplanques_2020}. Log-$F_0$ and voicing are obtained from the target recording using pYIN \cite{Mauch_2014_PyIN}. They therefore constitute oracle source information and are held fixed during comparisons.

The synthesis model is a six-layer Transformer encoder \cite{vaswani2017attention} with model dimension 384, conditioned on the speaker embedding and followed by a five-layer convolutional postnet. It predicts an 80-bin log-mel spectrogram at 50\,Hz. Waveforms are generated using
HiFi-GAN v1 \cite{kong2020hifigan}. The synthesizer is pretrained on the pseudo-labelled LibriTTS-R and subsequently fine-tuned on rtMRI-derived TVs.

The principal functional comparison uses the same held-out rtMRI utterances. Each utterance is synthesized twice, using inversion-derived TVs and corresponding rtMRI-derived TVs, with source features, speaker conditioning, and synthesis model held constant. The resulting difference isolates the acoustic effect of replacing inferred articulatory trajectories with independently rtMRI-derived trajectories.

Intelligibility is evaluated using word (WER) and character (CER) error rates from Whisper \cite{radford2023whisper}. Spectral distortion is measured using mel-cepstral distortion (MCD) \cite{Kubichek_1993_MCD}.

\subsection{Phonetic analysis of inversion errors}
\label{sec:phone_analysis}

To determine whether inversion errors are equally consequential across speech sounds, we compare the spectral output obtained from rtMRI-derived and inversion-derived TVs within each forced-aligned phone interval. For each token we compute the mean frame-level cepstral distance between predicted and ground-truth log-mel under each condition, over the full 80-bin mel range (0–8\,kHz), and report the difference (rtMRI-derived minus inversion-derived) as the spectral gap, holding source features and speaker conditioning fixed. 

Phones are grouped according to their characteristic constriction degree in Articulatory Phonology \cite{browman1992articulatory}: wide configurations (vowels), narrow non-turbulent configurations (approximants), closed configurations (stops and nasals), and critical constrictions (fricatives and affricates). The latter are produced with a highly constricted airway whose precise
geometry is important for turbulence generation.

Finally, because sibilant place provides a focused test of constriction location, we examine \textipa{/s/} and \textipa{/S/} \cite{jongman_acoustic_2000, forrest_statistical_1988}. We compare the spectral centroid of the frication noise between the reference and synthesized signals and quantify the separation between the two phone classes using Cohen's $d$.

\section{Results and Discussion}
\label{sec:results}

\subsection{How faithfully does inversion recover vocal-tract behaviour?}
\label{sec:results_fidelity}

\begin{table}[t]
\centering
\caption{Inversion fidelity on held-out speakers.}
\label{tab:tv_fidelity}
\setlength{\tabcolsep}{3pt}
\begin{tabular}{lcc@{\hspace{8pt}}lcc}
\toprule
TV & $r$ & $R_\sigma$ & TV & $r$ & $R_\sigma$ \\
\midrule
LA   & 0.720 & 0.848 & TBCD & 0.742 & 0.849 \\
LP   & 0.488 & 0.683 & TBCL & 0.340 & 0.595 \\
TTCD & 0.850 & 0.920 & TRCD & 0.794 & 0.831 \\
TTCL & 0.475 & 0.726 & VEL  & 0.737 & 0.747 \\
\midrule
\multicolumn{3}{c}{$\overline{r}=0.643$} &
\multicolumn{3}{c}{$\overline{R_\sigma}=0.775$} \\
\bottomrule
\end{tabular}
\vspace{-5pt}
\end{table}

Table~\ref{tab:tv_fidelity} shows that inversion accuracy varies substantially across articulatory dimensions. Constriction-degree variables are recovered most strongly, whereas constriction-location variables are considerably weaker. More importantly for physical interpretation, $R_\sigma$ is below unity for every TV, with a mean of 0.775 and a minimum of 0.595 for TBCL. The inferred trajectories therefore systematically occupy a narrower dynamic range than the rtMRI-derived trajectories, with the strongest compression occurring for the location variables. Thus, even when inferred and rtMRI-derived trajectories are strongly correlated, the inferred TVs do not necessarily preserve the full extent of rtMRI-derived articulatory movement.

\subsection{Do inversion errors matter acoustically?}
\label{sec:results_functional}

We next ask whether the differences between inferred and rtMRI-derived TVs are acoustically consequential. On the same held-out rtMRI utterances, inversion-derived TVs give 35.8\% WER, 23.1\% CER, and 3.77\,dB MCD, compared with 39.6\% WER, 26.2\% CER, and 3.86\,dB MCD when the same synthesizer is driven by rtMRI-derived TVs. Thus the aggregate effect of TV provenance is modest, motivating a more detailed analysis of whether the discrepancy is concentrated in particular articulatory configurations.

Three controls confirm that this comparison probes the articulatory representation rather than a failure elsewhere in the pipeline. In the pseudo-labelled LibriTTS-R domain, the same architecture reaches 7.7\% WER against a 5.4\% reference, so the synthesizer reconstructs speech accurately when driven by inversion-derived TVs in clean conditions. Ground-truth mel spectrograms passed through the vocoder yield 10.7\% WER against 10.3\% for the original-recording reference, so waveform generation contributes little to the observed error. Replacing the TVs with constants raises CER to 72.7\%, confirming that the articulatory input is load-bearing despite the oracle source features.

\subsection{Where do inversion errors become acoustically consequential?}
\label{sec:results_phone}

The modest aggregate difference between the two TV conditions conceals a strong phonetic structure. Table~\ref{tab:spectral_gap} groups phone tokens by their characteristic constriction degree. The spectral gap is 0.71\,dB for wide configurations, 0.40\,dB for narrow
approximants, and 0.75\,dB for closed stops and nasals. In contrast, fricatives and affricates requiring a critical constriction show a substantially larger gap of 2.47\,dB.

\begin{table}[t]
\centering
\caption{rtMRI-derived minus inversion-derived spectral gap, grouped by characteristic constriction degree in the sense of Articulatory Phonology \cite{browman1992articulatory}}
\label{tab:spectral_gap}
\begin{tabular}{llrr}
\toprule
Constriction & Classes & $n$ & Gap (dB) \\
\midrule
Wide & vowels & 2488 & 0.71 \\
Narrow & approximants & 369 & 0.40 \\
Closed & stops, nasals & 798 & 0.75 \\
Critical & fricatives, affricates & 558 & 2.47 \\
\bottomrule
\end{tabular}
\vspace{-5pt}
\end{table}

The pattern is not explained simply by constriction magnitude. Approximants also involve a narrow but non-zero constriction, yet their difference remains similar to that of vowels and complete closures. The spectral gap for critical constrictions is 2.07\,dB larger than for narrow approximants (2.47 vs. 0.40\,dB). What distinguishes fricatives and affricates is that a sufficiently narrow constriction must be maintained to generate turbulence, and the resulting spectrum depends strongly on both constriction geometry and the shape of the downstream cavity \cite{stevens_acoustic_2001}. The effect is predominantly carried by fricatives, with a gap of 2.25\,dB ($n=508$); affricates show a numerically larger gap, although only 50 tokens are available. This suggests that gross articulatory movements can be estimated sufficiently well for many speech sounds, but errors in fine constriction geometry become substantially more consequential when the acoustic output is highly sensitive to small articulatory changes.

\begin{figure}[t]
\centering
\includegraphics[width=\columnwidth]{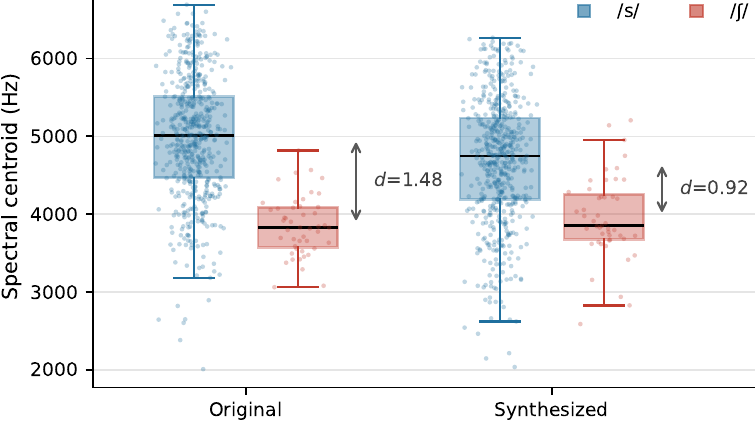}
\caption{Spectral centroid of frication for \textipa{/s/} and \textipa{/S/} in original and synthesized speech.}
\label{fig:sibilants}
\end{figure}

A focused \textipa{/s/}--\textipa{/S/} analysis provides a constriction-location example, shown in Fig.~\ref{fig:sibilants}. TTCL is among the least accurately recovered TVs, and the spectral-centroid separation between the two sibilants contracts from 1136 to 723\,Hz after synthesis, with Cohen's $d$ falling from 1.48 to 0.92 (62\% retained). This is consistent with loss of constriction-location information, although only 47 \textipa{/S/} tokens are available and the 16-kHz sampling rate limits analysis to 8\,kHz.

\section{Conclusions}

We used paired rtMRI and speech to test whether acoustically inferred vocal tract variables faithfully represent independently observed articulation. Fidelity is uneven: constriction degrees are recovered more strongly than locations, and inferred trajectories systematically underestimate the dynamic range of rtMRI-derived articulation. Using synthesis as a matched functional probe, we find these discrepancies are not equally consequential: the effect is concentrated in fricatives and affricates, where a critical constriction must be maintained, while a focused \textipa{/s/}--\textipa{/S/} analysis links the degradation to constriction location.

These results distinguish acoustic usefulness from physical fidelity: an inferred representation may support speech reconstruction without faithfully measuring fine vocal-tract geometry. Improving and independently validating constriction degree and location will be important if acoustic-to-articulatory inversion is to characterize the vocal-tract behaviour of individual speakers.

\section{Acknowledgments}
Funding: Australian Research Council Grant DP22010293.
\bibliographystyle{IEEEbib}
\bibliography{refs}

\end{document}